\documentclass[reprint,superscriptaddress,nofootinbib,amsmath,amssymb,prb,floatfix]{revtex4-1}

\usepackage[utf8]{inputenc}
\usepackage[T1]{fontenc} 
\usepackage{graphicx}
\usepackage{dcolumn}
\usepackage{bm}
\usepackage{braket}
\usepackage{color}
\usepackage{multirow}
\usepackage{amsmath}
\usepackage{amssymb}
\usepackage{bbm}
\usepackage{float}
\usepackage{epstopdf}

\begin{document}

\title{Vibronic Dephasing Model for Coherent-to-Incoherent Crossover in DNA}

\author{Patrick Karasch}
	\affiliation{Bremen Center for Computational Materials Science, Department of Physics, University of Bremen, Am Fallturm 1, 28359 Bremen, Germany}
 
\author{Dmitry A. Ryndyk}
	\email{dmitry.ryndyk@bccms.uni-bremen.de}
	\affiliation{Bremen Center for Computational Materials Science, Department of Physics, University of Bremen, Am Fallturm 1, 28359 Bremen, Germany}
	
\author{Thomas Frauenheim}
	\affiliation{Bremen Center for Computational Materials Science, Department of Physics, University of Bremen, Am Fallturm 1, 28359 Bremen, Germany}

\date{\today}

\begin{abstract}
  In this work we investigate the interplay between coherent and incoherent charge transport in
  cytosine-guanine (GC) rich DNA molecules. Our objective is to introduce a physically grounded
  approach to dephasing in large molecules and to understand the length dependent charge transport
  characteristics and especially the crossover from coherent tunneling to incoherent hopping regime
  at different temperatures. Therefore, we apply the vibronic dephasing model and compare the
  results to the B\"uttiker probe model which is commonly used to describe decoherence effects in
  charge transport. Using the full ladder model and simplified 1D model of DNA, we consider
  molecular junctions with alternating and stacked GC sequences and compare our results to recent
  experimental measurements.
\end{abstract}


\maketitle


\section{Introduction}

Scientific interest in charge transport characteristics of biological molecules, in particular
double-stranded DNA, has grown over recent decades because of its fundamental importance in live
sciences and potential for electronic applications. \cite{endres_colloquium_2004, Cuniberti2006}
Many experiments on charge transfer and transport have been realized showing a wide range of
contradictory outcomes covering insulating \cite{braun_dna-templated_1998, de_pablo_absence_2000,
  storm_insulating_2001, zhang_insulating_2002}, semiconducting \cite{porath_direct_2000}, and
conducting \cite{fink_electrical_1999, cai_self-assembled_2000, tran_charge_2000,
  rakitin_metallic_2001, yoo_electrical_2001, kasumov_proximity-induced_2001} behaviour. This large variety of
experimental results reflects the complexity of charge transfer processes in DNA and especially the
influence of environmental effects characterized by measurement conditions as well as the structural
conformation. At low temperatures the transport should be coherent, weakly influenced by interaction
of electrons with conducting environment. On the other hand, the main source of dephasing at finite
temperatures is elastic electron-vibron interaction due to large number of soft vibrations in
biological molecules.

Experimental as well as theoretical studies revealed two fundamental charge transport mechanisms in
DNA. Over short distances and at low temperatures (actually, as we will see, even room temperature
can be low enough) charge carriers move coherently through delocalized molecular orbitals. In the
case of the Fermi level position inside the conductance gap, this coherent tunneling regime is
represented by an exponentially length dependent transmission \cite{giese_direct_2001,
  jortner_charge_1998, tong_tunneling_2002, olofsson_electron_2001}. In the case of coherent
resonant transport, the conductance is oscillating function of the length because of interference
\cite{natallia_v._distance-dependent_2010}.

However, in the long-range regime at high temperatures charge carriers localize on the bases and
the main charge transfer occurs through incoherent hopping between them.  This can be modeled by
multistep hopping events between purine bases, i.e., bases with the lowest oxidation potentials and
shows a linear length dependence of transmission at high temperatures.\cite{jortner_charge_1998,
  meggers_sequence_1998, bixon_energetic_2000, kubar_efficient_2008}

The calculation of coherent charge transport is based on the Landauer-B\"uttiker approach (also
known as the scattering method), which is implemented usually with a help of the Green function
technique\cite{Datta95book,DiVentra08book,Ryndyk09inbook,Ryndyk2015}.

However, due to external influences and temperature dependent vibrations a coherent theory is often
not sufficient. In recent years there were some studies to expand the coherent theories to include
decoherence effects.

In the framework of the Green function approach dephasing resulting from environmental effects can
be accounted for using the \textit{B\"uttiker probe model}\cite{buttiker_role_1986,
  Buttiker1988, damato_conductance_1990, Nozaki08jpcc, Nozaki12prb, Chen14jpcl, Venkatramani14faraddis, Beratan15accchemres, Kilgour15jcp}. It is based on the idea of virtual probes coupled to the
system (see Sec.\,\ref{sec_BP}). This coupling controls the strength of the dephasing and is chosen
empirically.

More physical is the \textit{vibronic dephasing}, the contribution of which is especially important
at high temperatures. Vibrational effects can be taken into account through the corresponding
self-energy, which is dependent on the Green function at the same energy in the elastic
approximation\cite{bihary_dephasing_2005, cresti_electronic_2006,
  golizadeh-mojarad_nonequilibrium_2007, penazzi_self_2016} (see Sec.\,\ref{sec_VD}).

In this paper we consider local and elastic dephasing models, which are most important at low and
intermediate temperatures at low applied voltages, corresponding to typical experimental conditions.

The recent experiment\cite{xiang_intermediate_2015} has shown intermediate coherent-incoherent
charge transport characteristics at room temperature in GC-rich DNAs. The phenomenological
explanation based on the incoherent hopping approach with addition of coherent transport was
suggested\cite{xiang_intermediate_2015}. It is important to understand transition from coherent to
incoherent transport starting from unified microscopic approach. The model of such crossover was
studied recently\cite{kim_intermediate_2016} by using B\"uttiker probe model of environmental caused
dephasing. In this paper we will investigate decoherence effects in DNA structures based on the
vibronic dephasing model proposed by some of us earlier\cite{penazzi_self_2016}. Besides, we
implemented both methods in the open software package {\textsc{DFTB$^{\text{+}}$XT\ }}\cite{DFTBXT}.

The paper is organized as follows. In Sec.\,\ref{sec_deph} we consider the Green function method and
elastic dephasing models. In Sec.\,\ref{sec_DNA} we introduce the effective tight-binding like
models of DNA and consider the particular system Hamiltonians and parameters. The results of
calculations and discussion are presented in Sec.\,\ref{sec_results}. And finally we give
conclusions and outlook in Sec.\,\ref{sec_conclusion}.

\section{Elastic dephasing}
\label{sec_deph}

In this section we outline the nonequilibrium Green function formalism for systems with elastic
dephasing. The total Hamiltonian in the case of two equilibrium electrodes can be written as
\begin{align}
    \boldsymbol{\hat{H}} =  \boldsymbol{\hat{H}_\text{L}} + \boldsymbol{\hat{H}_\text{LM}} + \boldsymbol{\hat{H}_\text{M}} + \boldsymbol{\hat{H}_\text{MR}} + \boldsymbol{\hat{H}_\text{R}},
\end{align}
where $\boldsymbol{\hat{H}_\text{L/R}}$ describes the left/right lead and
$\boldsymbol{\hat{H}_\text{M}}$ represents the DNA molecule. The coupling between leads and molecule
is defined by $\boldsymbol{\hat{H}_\text{LM}}$ and $\boldsymbol{\hat{H}_\text{MR}}$. The
corresponding retarded and advanced Green functions of the molecule are then given by
\begin{align} 
    \textbf{G}^\text{R} & (\varepsilon) = 
    \left[\boldsymbol{\mathbbm{1}}\varepsilon-\boldsymbol{\hat{H}_\text{M}}-
      \boldsymbol{\Sigma}^\text{R}_{\text{L}}(\varepsilon)-\boldsymbol{\Sigma}^\text{R}_{\text{R}}(\varepsilon)
      -\boldsymbol{\Sigma}^\text{R}_{\text{deph}}(\varepsilon)\right]^{-1}, \\
    \textbf{G}^\text{A} & (\varepsilon) = (\textbf{G}^\text{R}(\varepsilon))^\dagger,
\end{align}
where $\boldsymbol{\Sigma}^\text{R}_{\text{L/R}}(\varepsilon)$ is the retarded self-energy of the
left/right electrode and $\boldsymbol{\Sigma}^\text{R}_{\text{deph}}(\varepsilon)$ is the
self-energy due to dephasing. In the case of {\it elastic} dephasing, considered here, the
self-energy at some energy does not depend on other energies, but can be a function of the Green
function at the same energy. We do not take into account electron-electron interactions. In the case of dephasing in strongly interacting many-body systems other approaches may be required \cite{Esposito10jpcc,White13prb}.
  
In order to determine the transport properties, we use the many-body nonequilibrium Green function
formalism. The details can be found in Refs.\cite{Ryndyk09inbook,Ryndyk2015}. Within this approach
the current can than be calculated via the Meir-Wingreen formula
\begin{align}\label{MWJ}  \notag
    I = \frac{\mathrm{i}e}{h} \int^\infty_{-\infty} \text{Tr}\Big[\big(\boldsymbol{\Gamma}_\text{L}(\varepsilon)-\boldsymbol{\Gamma}_\text{R}(\varepsilon)\big)\textbf{G}^\text{<}(\varepsilon)\\ 
    +\big(f_\text{L}{(\varepsilon)}{\boldsymbol{\Gamma}_\text{L}(\varepsilon)} - f_\text{R}{(\varepsilon)}{\boldsymbol{\Gamma}_\text{R}(\varepsilon)}\big) 
    \big(\textbf{G}^\text{R}(\varepsilon) - \textbf{G}^\text{A}(\varepsilon)\big)\Big]
     d\varepsilon,
\end{align}
where the broadening function can be determined by:
$\boldsymbol{\Gamma}_\text{L/R}(\varepsilon) = \text{i} (\boldsymbol{\Sigma}_\text{L/R}(\varepsilon) - \boldsymbol{\Sigma}_\text{L/R}^\dagger(\varepsilon))$. \\
The lesser/greater green functions are defined through the Keldysh equation
\begin{align}  
    \textbf{G}^\gtrless(\varepsilon) = \textbf{G}^\text{R}(\varepsilon)\boldsymbol{\Sigma}^\gtrless_\text{tot}(\varepsilon)
    \textbf{G}^\text{A}(\varepsilon),\
\end{align}
where the total self-energy
$\boldsymbol{\Sigma}^\gtrless_\text{tot}(\varepsilon) =
\boldsymbol{\Sigma}^\gtrless_{\text{L}}(\varepsilon) +
\boldsymbol{\Sigma}^\gtrless_{\text{R}}(\varepsilon) +
\boldsymbol{\Sigma}^\gtrless_{\text{deph}}(\varepsilon)$ contains contributions from the leads as
well as from interactions. The former can be expressed by the equilibrium fermi functions of the
leads $f_\text{L/R}(\varepsilon)$ via
\begin{align}
  &\boldsymbol{\Sigma}^\text{<}_\text{L/R}(\varepsilon) = \text{i} f_\text{L/R}(\varepsilon)\boldsymbol{\Gamma}_\text{L/R}(\varepsilon), \\ 
  &\boldsymbol{\Sigma}^\text{>}_\text{L/R}(\varepsilon) = -\text{i}\left(1-f_\text{L/R}(\varepsilon)\right)\boldsymbol{\Gamma}_\text{L/R}(\varepsilon) \,.
\end{align}
By assuming that the total self-energy $\boldsymbol{\Sigma}_\text{tot}$ contains just contributions
of the leads, eq. (\ref{MWJ}) would yield the Landauer formula for coherent transport.

\subsection{Vibronic dephasing (VD) model}
\label{sec_VD}

Incoherence effects on charge transport can be taken into account by including the electron-vibron
interaction. This can be realized by an additional self-energy $\boldsymbol{\Sigma}_\text{vib}$.

In order to obtain an expression for this self-energy we assume that the Hamiltonian of the central
device can be written as:
\begin{align}
  \boldsymbol{\hat{H}_\text{M}} = \boldsymbol{\hat{H}_\text{el}} + \boldsymbol{\hat{H}_\text{e-v}}.
\end{align}
The electronic structure of the molecule is given by $H_\text{el}$ and the electron-vibron
interaction $H_\text{e-v}$ can be written as:
\begin{align}
  \boldsymbol{\hat{H}_\text{e-v}} = \sum_{ij,\alpha}M_{ij}^\alpha (a_\alpha + a_\alpha^\dagger)d_i^\dagger d_j,
\end{align}
where $a(a^\dagger)$ describes the annihilation (creation) operators of vibrons and $d(d^\dagger)$
the same for electrons. The coupling between electrons and vibrons in mode $\alpha$ is described by
$M^\alpha$.  The corresponding self-energy reads (see details in Ref.\cite{Ryndyk2015}, actually it
is enough to calculate the lesser self-energy)
\begin{align}
  \label{sigma_r}
\mathbf{\Sigma}^{<}_\text{vib}(\varepsilon) = 
\sum_\alpha \frac{i}{2\pi} \int \mathbf{M}^\alpha\mathbf{G}^<(\varepsilon - \omega)\mathbf{M}^\alpha \mathbf{D}^<_{0,\alpha}(\omega) d\omega,
\end{align}
where $\mathbf{D}_{0,\alpha}$ is the free-vibron Green function.

To study the decoherence effects at low voltages (linear conductance) we use the elastic dephasing
model \cite{penazzi_self_2016,bihary_dephasing_2005,golizadeh-mojarad_nonequilibrium_2007,
  cresti_electronic_2006}
which will be refered to as \textit{Vibronic Dephasing (VD)}.

Within this model it is assumed that the electron-vibron interaction is localized at atomic sites
$\alpha = i$ and is identical at all sites.  Moreover, the energy of vibronic quanta
$\hbar\omega_\alpha$ is assumed to be small compared to the other energy scales in the system, so
that no vibrons are excited inelastically. This approximation works well for low-frequency acoustic
type vibrations.

In the elastic approximation one can integrate (\ref{sigma_r}) over $\omega$ assuming $\mathbf{G}^<(\varepsilon - \omega) \approx\mathbf{G}^<(\varepsilon)$. 
The retarded and lesser self-energies are then represented by:
\begin{align}\label{sigma_vd}
\left[\mathbf{\Sigma}_\text{VD}^{R(<)}(\varepsilon)\right]_{ij} = \gamma^2 \mathbf{G}^{R(<)}_{ii}(\varepsilon) \delta_{ij},
\end{align}
where $\gamma$ is the strength of dephasing. It can be estimated as \cite{penazzi_self_2016,
  bihary_dephasing_2005}
\begin{align}\label{gamma_vd}
 \gamma \approx \sqrt{k_\text{B}TV_\text{e-v}},
\end{align}
where $V_\text{e-v}$ describes the electron-vibron-coupling and $T$ is the temperature.

Since the self-energy $\mathbf{\Sigma}_\text{VD}^{R(<)}(\varepsilon)$ depends explicitly on
$\mathbf{G}^{R(<)}(\varepsilon)$ we have to solve this problem self-consistently.

\subsection{B\"uttiker probe (BP) model}
\label{sec_BP}

\begin{figure}[b]
   \includegraphics[width=0.99\columnwidth]{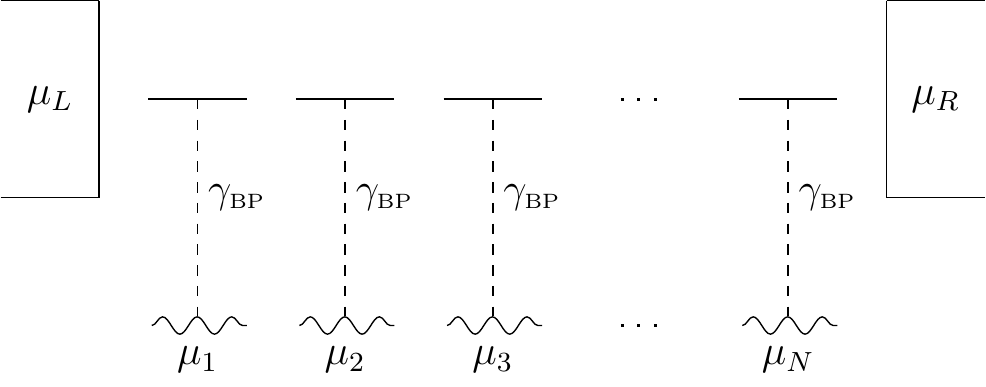}
   \caption{Schematic representation of the B\"uttiker probe model. Virtual leads with chemical potentials $\mu_n$ are coupled to every site of the finite tight binding chain to introduce dephasing. \label{BPPic}}
\end{figure}

Another way to include dephasing in charge transport calculations is the so called
\textit{B\"uttiker Probe (BP) model} \cite{buttiker_role_1986, Buttiker1988,
  damato_conductance_1990, Nozaki08jpcc, Nozaki12prb, Chen14jpcl, Venkatramani14faraddis,
  Beratan15accchemres, Kilgour15jcp}.  As depicted in Fig.\,\ref{BPPic}, in this model virtual probes
are connected to every site of the chain.

This leads to an additional self-energy (in the simplest case site-independent)
\begin{align}
 \boldsymbol{\Sigma^\text{R}_\text{BP}} = \text{-i}\frac{\gamma_\text{BP}}{2},
\end{align}
where $\gamma_\text{BP}$ is the coupling strength of the probes to the sites. Electrons may tunnel
to the probes and back instead of travelling from left to right lead directly. To describe model
dephasing, in contrast to the real conducting environment, the probes have to fulfill a zero net
current condition i.e. the number of electrons entering the probe has to be the same as the number
of electrons leaving it again to the same site.  The effective transmission between left and right
leads is then given by \cite{damato_conductance_1990}
\begin{align}
  T_\text{eff} = T_\text{LR} + T_\text{corr}
\end{align}
where $T_\text{LR}$ is the coherent transmission from left to right lead.

The second term includes the transmission corrections due to the probes
\begin{align}\label{Tcorr}
 T_\text{corr} = \sum_{m,n}^N T_{\text{L}m} W^{-1}_{mn} T_{n\text{R}}
\end{align}
where $W$ describes the Markov matrix:
\begin{align}
  W_{mn} = (1-R_m)\delta_{mn} - (1-\delta_{mn})T_{mn}\,.
\end{align}
The reflection function is defined as:
\begin{align}
    R_m = 1-\sum_{n\neq m}^N T_{mn}.  
\end{align}
where $T_{ij}$ describes all transmissions between probes $i,j=m,n$ and two leads $i,j=L,R$:
\begin{align}\label{GGGG}
  T_{ij} = \text{Tr}\left(\boldsymbol{\Gamma}_i \textbf{G}^\text{R} \boldsymbol{\Gamma}_j \textbf{G}^\text{A} \right) \, .
\end{align}
This approximation is within the Landauer-B\"uttiker picture and we can use the coherent expression
(\ref{GGGG}) for transmissions.

This model introduces decoherence effects due to re-emission of electrons with random phase. This
means that the coupling strength $\gamma_\text{BP}$ of the probes controls the dephasing
strength. The transmission correction in eq. \eqref{Tcorr} is only valid for small voltages and
small temperatures. Otherwise a numerical approach to adjust the chemical potential $\mu_n$ of the
probes should be used to ensure zero net
current.\cite{golizadeh-mojarad_nonequilibrium_2007,kim_intermediate_2016,Korol1712.08515}

Although the vibronic dephasing model and the B\"uttiker probe model have different origins of
dephasing, both describe elastic scattering caused by local interactions.

\section{The model of DNA}
\label{sec_DNA}

\begin{figure*}
 \includegraphics[width=0.95\textwidth]{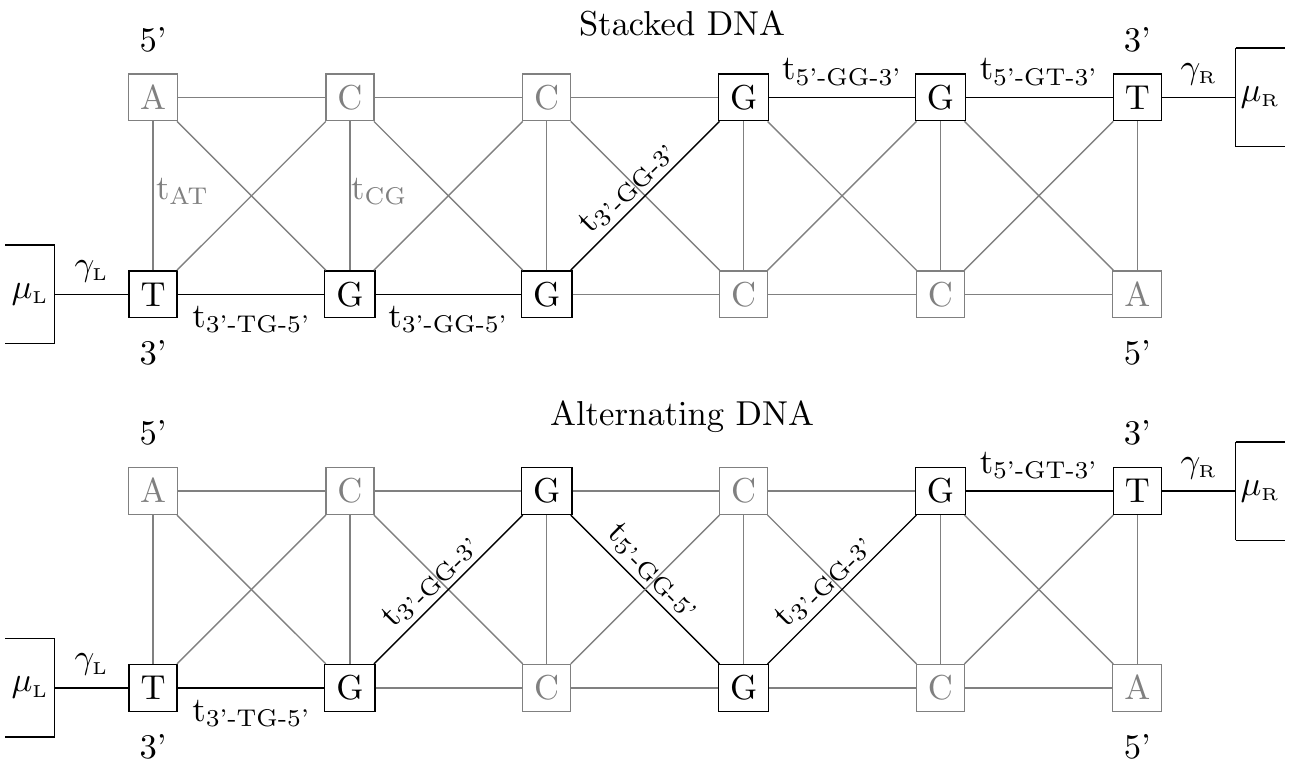}
 \caption{Schematic representation of the ladder models for $n=2$ (gray) and the corresponding 1D models (black) with hopping parameters $t$. Onsite energies and charge transfer integrals are taken from Ref.\citenum{senthilkumar_absolute_2005}.}
 \label{Models}
\end{figure*}

Our aim is to analyze the interplay between coherent and incoherent charge transport in DNA.
Therefore, we will investigate stacked $\text{AC}_n\text{G}_n\text{T}$ and alternating $\text{A(CG)}_n\text{T}$ sequences following Xiang et al.\cite{xiang_intermediate_2015}, where A,C,G,T stand for the nucleobases adenine, cytosine, guanine and thymine respectively. These structures are double-stranded and self-complementary.\\
We follow the idea of previous authors\cite{yi_conduction_2003, gutierrez_inelastic_2006,
  wang_charge_2006, zilly_conductance_2010, kim_intermediate_2016,Korol1712.08515} and model these
structures by using a next neighbor tight-binding (TB) ladder models as depicted in
Fig.~\ref{Models} (gray lines). Within this approach we neglect any geometrical effects and assume
that they are included in the electronic parameters. On top of that we consider low-energy charge
transport so that just the frontier orbitals of the bases are relevant. This means that we take one
orbital per base into account, i.e., we have $N=4n+4$ electronic states per molecule with $n$ base
pairs.

The corresponding Hamiltonian reads:
\begin{align}
\boldsymbol{\hat{H}} = \sum_{i=1}^{n}  \left(\sum_{s=1,2} \varepsilon_{i,s} \hat{c}_{i,s}^\dagger \hat{c}_{i,s} + \sum_{s\neq s'=1,2} t_{i,s,s'} \hat{c}_{i,s}^\dagger \hat{c}_{i,s'} \right. \notag\\ 
 \left. + \sum_{s,s' = 1,2} t_{i,i+1,s,s'} (\hat{c}_{i,s}^\dagger c_{i+1,s'} + h.c.)  \right)
\end{align}
where $\varepsilon_{i,s}$ describes the onsite energy of site $i$ in strand $s$, $t_{i,s,s'}$ the
interstrand hopping within a base pair and $t_{i,i+1,s,s'}$ charge transfer between the base pairs
(interstrand as well as intrastrand).

Furthermore we adopt the beforehand mentioned assumption that charge transport occurs only through
the purine bases (T, G).

We compare the ladder models with the corresponding 1D models which just include these bases. For a
visualisation see the black lines in Fig.~\ref{Models}.

We use the electronic parameters which were calculated on a DFT level by Senthilkumar et
al.\cite{senthilkumar_absolute_2005} and modify the data set to reproduce experimental results from
Xiang et al. \cite{xiang_intermediate_2015}.  A short explanation of this adjustment is given in the
appendix.  While most of the charge transfer integrals can be used directly we have to pay some
attention to the onsite energies. In the study these values were estimated depending on the two
neighboring bases and show, e.g.  for guanine, a maximal variation of
$\Delta \varepsilon_\text{G} \approx 0.5\, \text{eV}$. Since the outer bases, i.e., A and T are not
sandwiched between two bases, we follow the idea of previous studies\cite{zilly_conductance_2010,
  kim_intermediate_2016} and average the onsite energies.  The numerical values of the parameter set
can be found in the appendix.

We investigate a single double-stranded DNA sandwiched between two leads.  The influence of the
leads will be modelled in the wideband limit, i.e. we neglect any energy dependence of the coupling.
Furthermore we assume that contacts are just coupled to the first and last base (thymine) of the DNA
molecule.  This assumption is in fair agreement with the experiment of Xiang et
al. \cite{xiang_intermediate_2015}.  The hybridization functions in Eq.\,(\ref{MWJ}) than reduces to
\begin{align}
 &\left[\boldsymbol{\Gamma}_\text{L}\right]_{1,1} = \gamma_\text{L}, \\
 &\left[\boldsymbol{\Gamma}_\text{R}\right]_{N,N} = \gamma_\text{R}\,.
\end{align}
Therefore the coupling to the leads is controlled by the adjustable parameters $\gamma_\text{L/R}$.

With these parameters we can calculate the transport properties for the ladder as well as for the 1D
models.

\section{Results}
\label{sec_results}

\subsection{Computational Details}
\label{subsec_comp}

Our calculations were performed using {\textsc{DFTB$^{\text{+}}$XT\ }} package \cite{DFTBXT}, based
on the {\textsc{DFTB$^{\text{+}}$}} \cite{Aradi07jpca,Pecchia08njp} source code. The calculation of
the vibrational self-energies (\ref{sigma_vd}) is done self-consistently.

The coupling to the electrodes is estimated to be $\gamma_\text{L/R} = 0.30\,\text{eV}$. This value is typical for the couplings of molecular orbitals of small organic molecules to gold surfaces \cite{Ryndyk13prb}.

The vibronic dephasing parameter we estimate as
$\gamma_\text{VD} = \sqrt{0.03\,\text{eV}^2/\text{K}\cdot\text{T}}$ from the formula
(\ref{gamma_vd}). The voltage is taken to be $0.002\,\text{V}$ to be in the linear regime. The Fermi
energy of the leads are aligned with the HOMO energy $\varepsilon_G$, which is the standard
approximation explained by the level
alignment\cite{kim_intermediate_2016,natallia_v._distance-dependent_2010}.

\subsection{Comparison of Ladder and 1D DNA models}

\begin{figure}[t]
 \includegraphics[width=0.99\columnwidth]{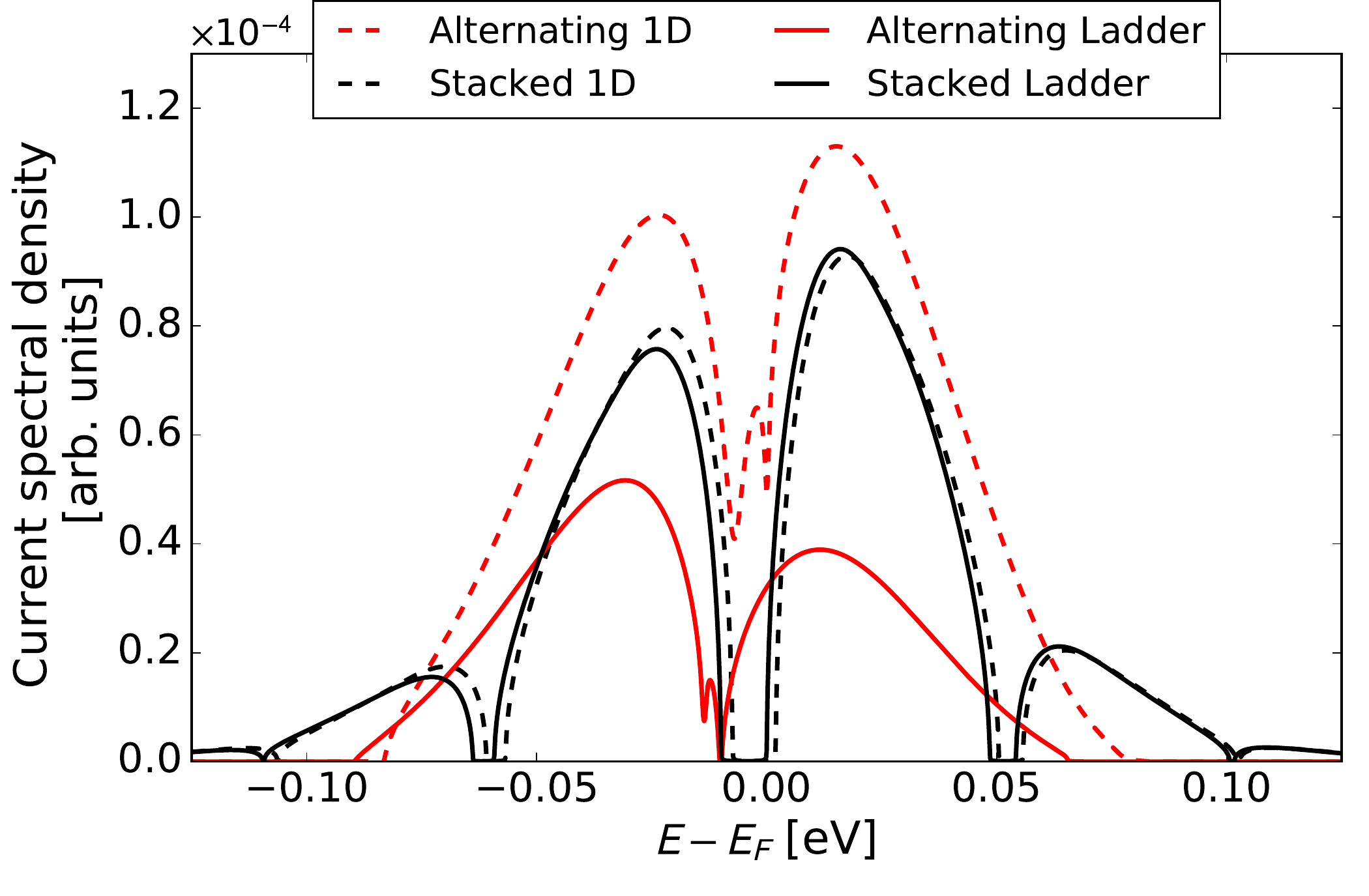}
  \caption{Current spectral densities of the DNA ladder models (solid lines) and the corresponding 1D models (dashed lines) for $n = 8$ at room temperature ($\gamma_\text{VD} = 0.03\,\text{eV}$)}
 \label{Tunneling_Models}
\end{figure}

In order to get insight into the charge transport characteristics of the two different TB models we
calculate the current spectral density which is the integrand of Eq.~\eqref{MWJ}.  The results for
the stacked and alternating DNA molecules with 8 base bairs with dephasing are shown in
Fig.~\ref{Tunneling_Models}. The completely coherent current spectral densities are equivalent to
the transmission functions from Landauer-B\"uttiker theory \cite{Ryndyk2015, penazzi_self_2016}.

The current spectral densities in Fig.~\ref{Tunneling_Models} of the stacked ladder and the
corresponding 1D model show good agreement, i.e., charge transport mainly occurs through the
purine bases.  Contrary the alternating models show a large discrepancy. The current spectral
density of the ladder model is shifted to lower energies and suppressed by a factor of
$\approx 2$ corresponding to the density of the 1D model. \\
This large difference is caused by the strong intrastrand coupling
$t_{5'-\text{GC}-3'} = t_{3'-\text{CG}-5'}$ in the alternating model which is neglected in the 1D
model. This has a stronger influence on the alternating than on the stacked DNA because it appears
just once in the latter system. Nevertheless the current spectral density of the 1D alternating DNA
model shows the same features as the ladder model, i.e. it can be used to investigate qualitative
transport properties. It should be emphasized that the 1D alternating model will lead to a higher
current and thus a lower resistance than the stacked model. Therefore we will investigate dephasing
effects utilizing the ladder model to get comparable results.

\subsection{Dephasing in the Ladder models}
 
\begin{figure*}
  \includegraphics[width = 0.99\textwidth]{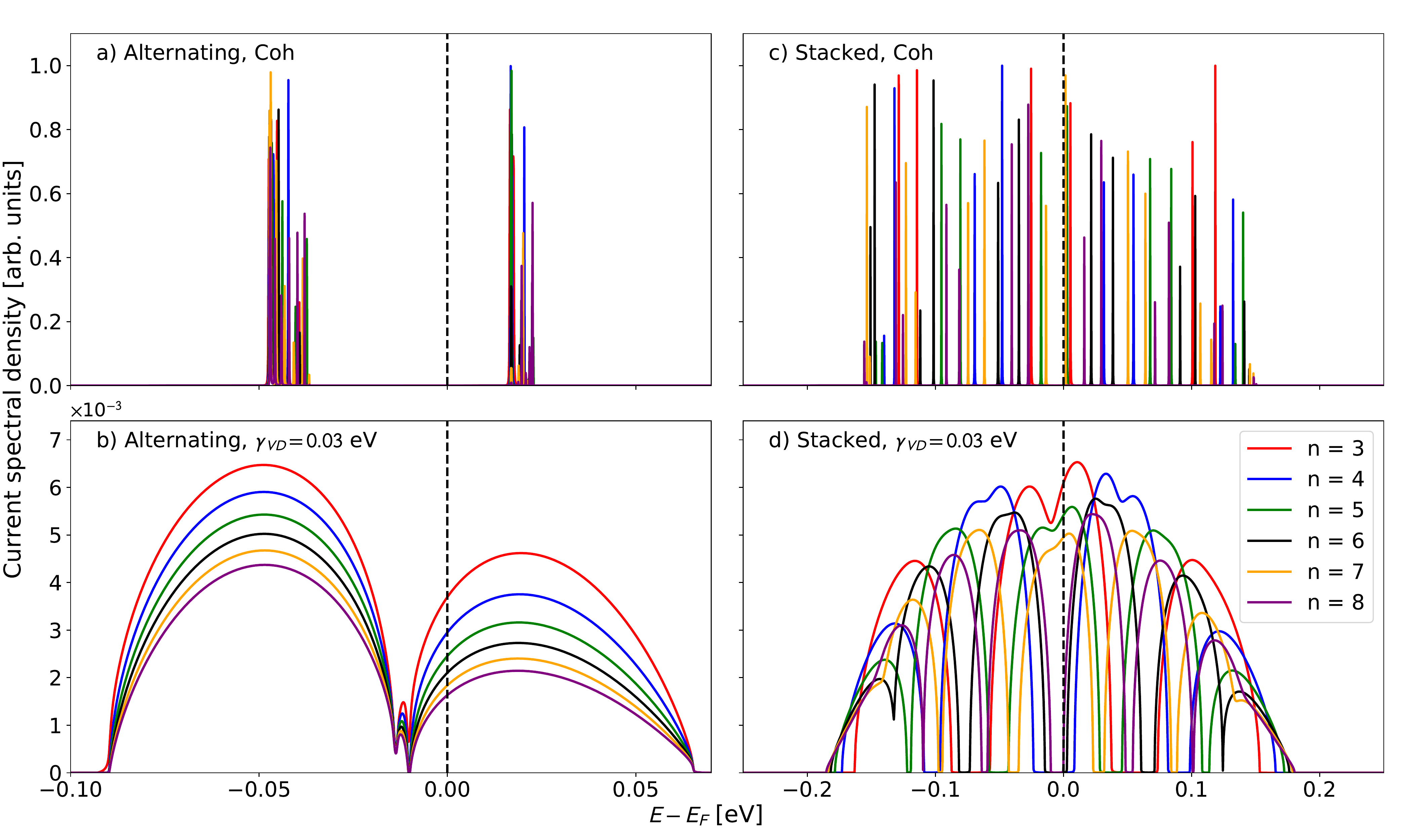}
  \caption{Current spectral densities calculated using the ladder models at room temperature ($T = 295\,\text{K}$).
Left panel: Alternating A(CG)$_n$T DNA for a) coherent case and b) case with dephasing. 
Right panel: Stacked AC$_n$G$_n$T c) coherent case and d) case with dephasing.
Dashed lines indicated the position of the Fermi energy of the leads $E_F = \varepsilon_G$}
  \label{Transmissions}
\end{figure*}

We apply the vibronic dephasing model and the ladder models described before to investigate the
effect of dephasing. Therefore we calculate the current spectral densities for both DNA molecules
for $n=3,...,8$ base pairs at room temperature. The results for the coherent and incoherent cases
are shown in Fig.~\ref{Transmissions}.  The two DNA molecules show quite narrow transmission
windows. For the coherent calculations (upper pannel in Fig.~\ref{Transmissions}) we find gaps at
the Fermi energy of the leads $E_F = \varepsilon_G$, which is in agreement with experimental results
from Shapir et al.\cite{shapir_electronic_2007,Ryndyk09acsnano}. The transmission functions of the 6 alternating DNA
molecules in Fig.~\ref{Transmissions}a are grouped in two discrete transmission windows above and
below the Fermi energy of the leads. The stacked molecules (Fig.~\ref{Transmissions}c) show broad
distributed transmission peaks. Furthermore the molecules with an odd number of base pairs show
transmission peaks very near $E_F$ while in the alternating molecules there are no features near
$E_F$.

Including decoherence effects leads to broadening of the current spectral densities of both DNA
molecules. Moreover it can be seen that this broadening leads to the disappearance of the
transmission gap at $E_G$ for alternating DNA (Fig.~\ref{Transmissions}b) and the appearance of a
local maximum slightly below $E_F$.  Additionally it can be seen that the current spectral densities
of all alternating molecules have the same shape where the maxima decrease monotonously with the
system length.

The incoherent current spectral densities for the stacked DNA molecules 
(Fig.~\ref{Transmissions}d) show two different shapes. For the molecules with an 
odd number of base pairs the transmission gap is closed in contrast to the 
molecules with even number of base pairs. This gap decreases with larger system 
sizes. It is also noticeable that all stacked curves shows additional gaps apart 
$E_F$. For example for $n=7$ (orange lines in Fig.~\ref{Transmissions}d) we can 
find transmission gaps at $-0.097$\,eV, $-0.039$\,eV, $0.028$\,eV and 
$0.086$\,eV relative to the Fermi energy. 

\begin{figure}[b]
  \includegraphics[width = 0.99\columnwidth]{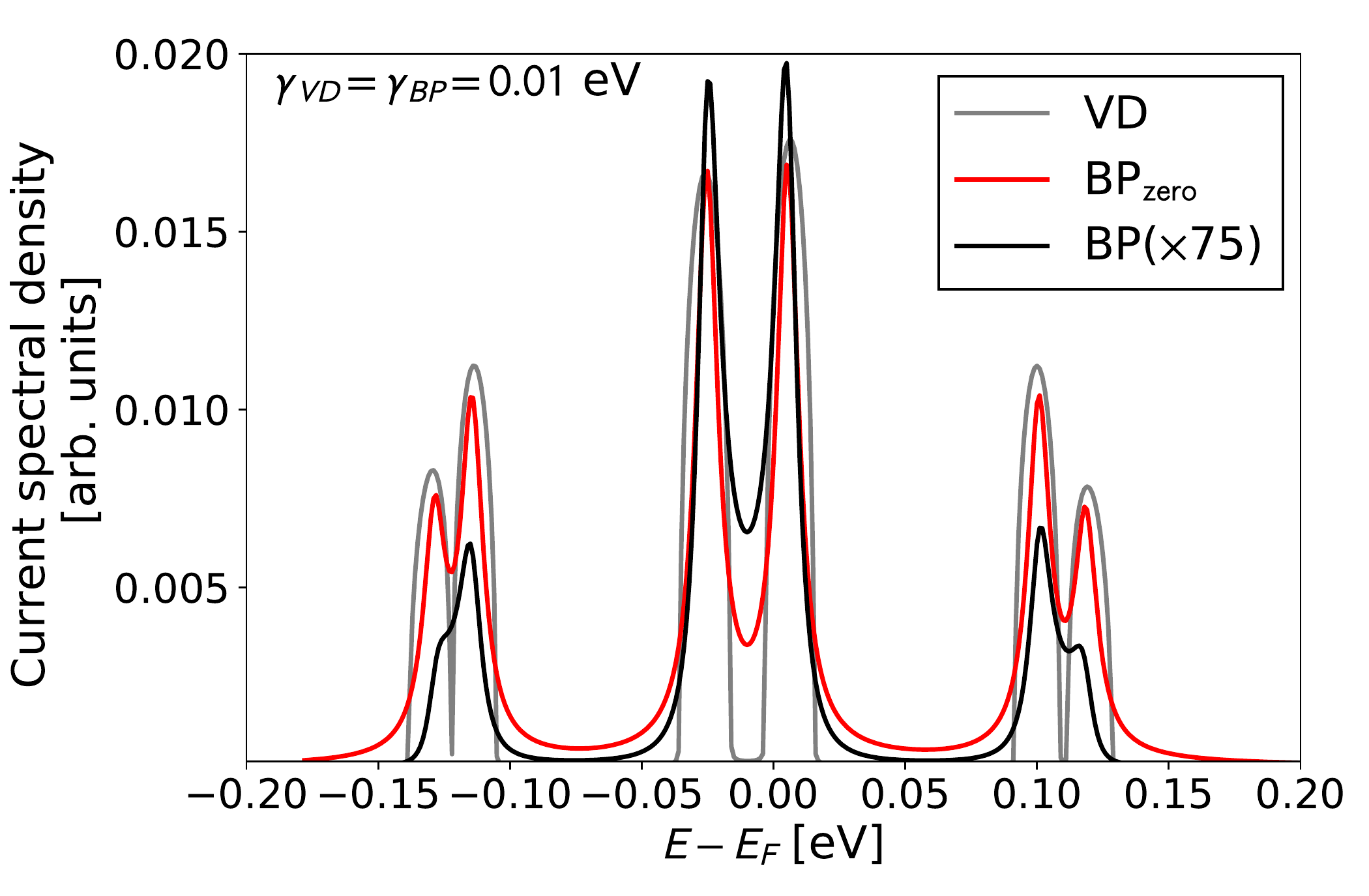}
  \caption{Current spectral densities calculated using the stacked ladder model 
for n = 3 at room temperature ($T = 295\,\text{K}$) for vibronic 
dephasing model (grey) and B\"uttiker Probe model with (red) and without 
  (black) zero current condition. For better visualization the 
B\"uttiker Probe model without zero current condition was enhanced.}
  \label{Transmissions_models}
\end{figure} 

To compare the vibronic dephasing model to the B\"uttiker probe model we calculated the current
spectral densities for a stacked DNA molecule with $n=3$ base pairs (see
Fig. \ref{Transmissions_models}). We have chosen a dephasing strength
$\gamma_{\text{VD}} = \gamma_{\text{BP}} = 0.01\,\text{eV}$ which is smaller than in
Fig.\,\ref{Transmissions} to emphasize the difference between the dephasing models. The peak
positions are identical for the three models but the shape differs. While the peaks of the vibronic
dephasing model (grey lines in Fig. \ref{Transmissions_models}) are rather sharp the BP models show
a more smeared shape.  Furthermore the minima are not that pronounced in the BP models as in the VD
model leading to the disappearance of the gap at Fermi energy.

Additionally the BP model without zero net current condition (black lines) shows a much smaller
current spectral density caused by leaking currents through the probes. This model may be important
and corresponds to the other physical case, namely the conducting environment.

\subsection{Length-dependent resistance}

\begin{figure}[b]
  \includegraphics[width = 0.99\columnwidth]{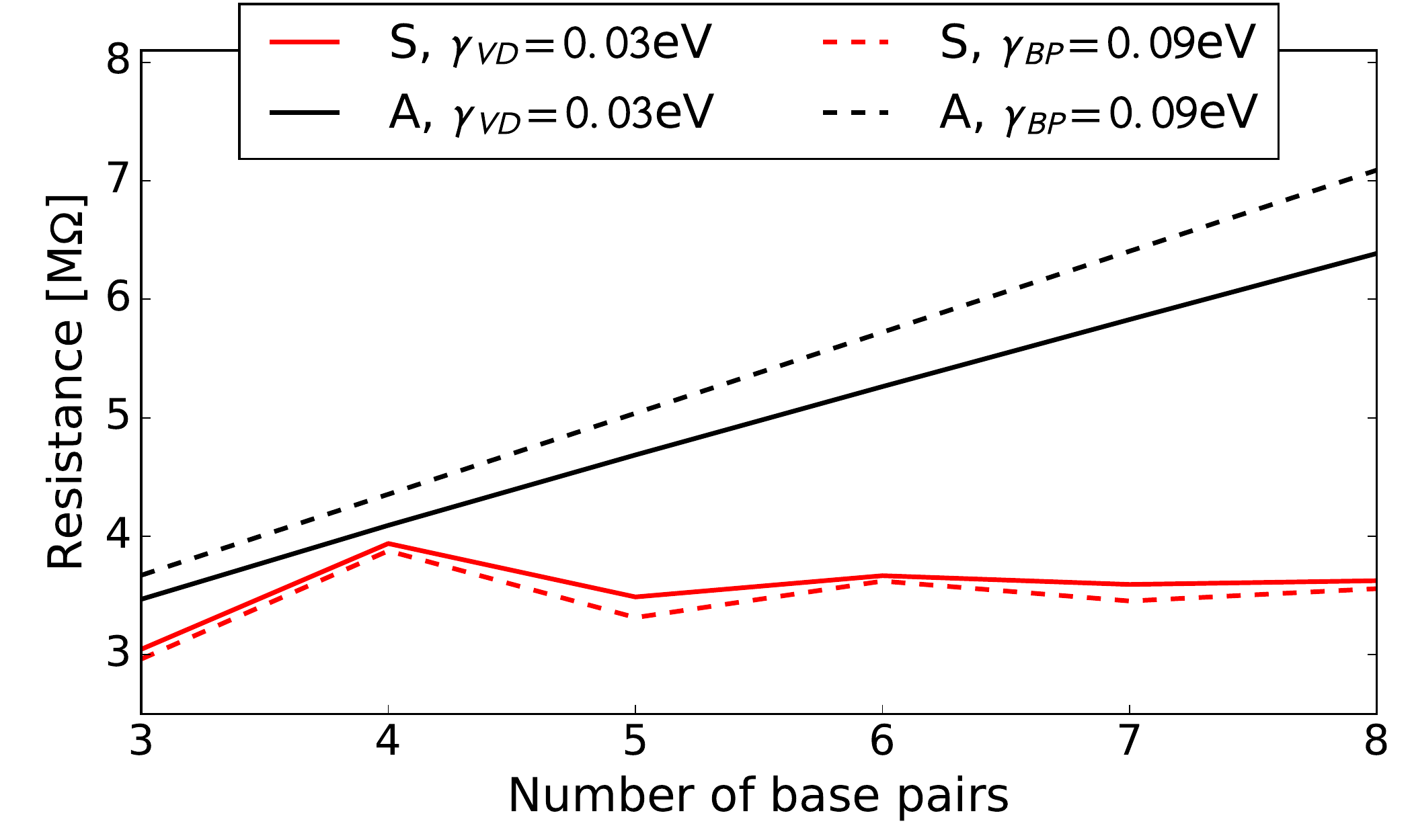}
  \caption{Length dependent resistance for alternating and stacked DNA, 
calculated with ladder model for vibronic dephasing model (solid lines) as well 
as B\"uttiker probe model (dashed lines).}
 \label{R_N}
\end{figure}

With the previously discussed current spectral densities we can calculate the current through the
molecules according to Eq. \eqref{MWJ} and thus the resistance. We used a vibronic dephasing
strength of $\gamma_\text{VD} = 0.03\,\text{eV}$ as in Fig. \ref{Transmissions} and estimated
$\gamma_{\text{BP}} = 0.09\,\text{eV}$ to reproduce a similar behavior in the length dependence.
The resistance at room temperature for the VD model (solid lines) as well as for the BP model with
zero net current condition (dashed lines) is shown in Fig.~\ref{R_N}.

The resistance of the stacked DNA molecule shows overall a slight slope and additionally an
even-odd-oscillation. This oscillations are slightly more pronounced by the BP model than by VD
model.  It is well known that these oscillations are caused by a partial delocalization of charge
carriers in the stacked G-sequences. \cite{kim_intermediate_2016}.  This can also be seen in
Fig.~\ref{Transmissions}d: The current spectral densities of molecules with odd numbers show a
maximum at $E_F$ and thus a minimum in resistance. 

In comparison the resistance of the alternating molecule just shows a linear 
length dependence. The VD model has a slope of 
$0.584$\,M$\Omega$ per base pair  similar to the 
experiments\cite{xiang_intermediate_2015} ($\approx 0.596$\,M$\Omega$ per
base pair). In the BP model the slope is slightly steeper with
$0.684$\,M$\Omega$ per base pair. 
This increase can also be understood with the current spectral density in 
Fig.~\ref{Transmissions}b): In the incoherent case a length dependent decrease 
of the maxima can be seen. This leads to a lowered current and thus to a higher 
resistance.
Length-dependent resistance measurements of Xiang et al. \cite{xiang_intermediate_2015} show also a linear increase for alternating  and stacked DNA and oscillations which superimpose the 
linear increase in the stacked molecule.  However these measurements show a similar slope of the $R(N)$ curves for stacked and alternating DNA sequences which we cannot reproduce with our model calculations.

In a previous study of Kim et al. \cite{kim_intermediate_2016}, where decoherence effects caused by
environmental effects were investigated using the B\"uttiker Probe model, a similar difference
between alternating and stacked DNA molecules as in our model calculations was found. Note, that in
Ref.\,\onlinecite{kim_intermediate_2016} the probes were considered at finite temperature, and the
zero-current conditions for probes with a certain broadening for the derivative of the Fermi
function were used.

\subsection{Temperature-dependent resistance}

\begin{figure}[t]
  \includegraphics[width = 0.99\columnwidth]{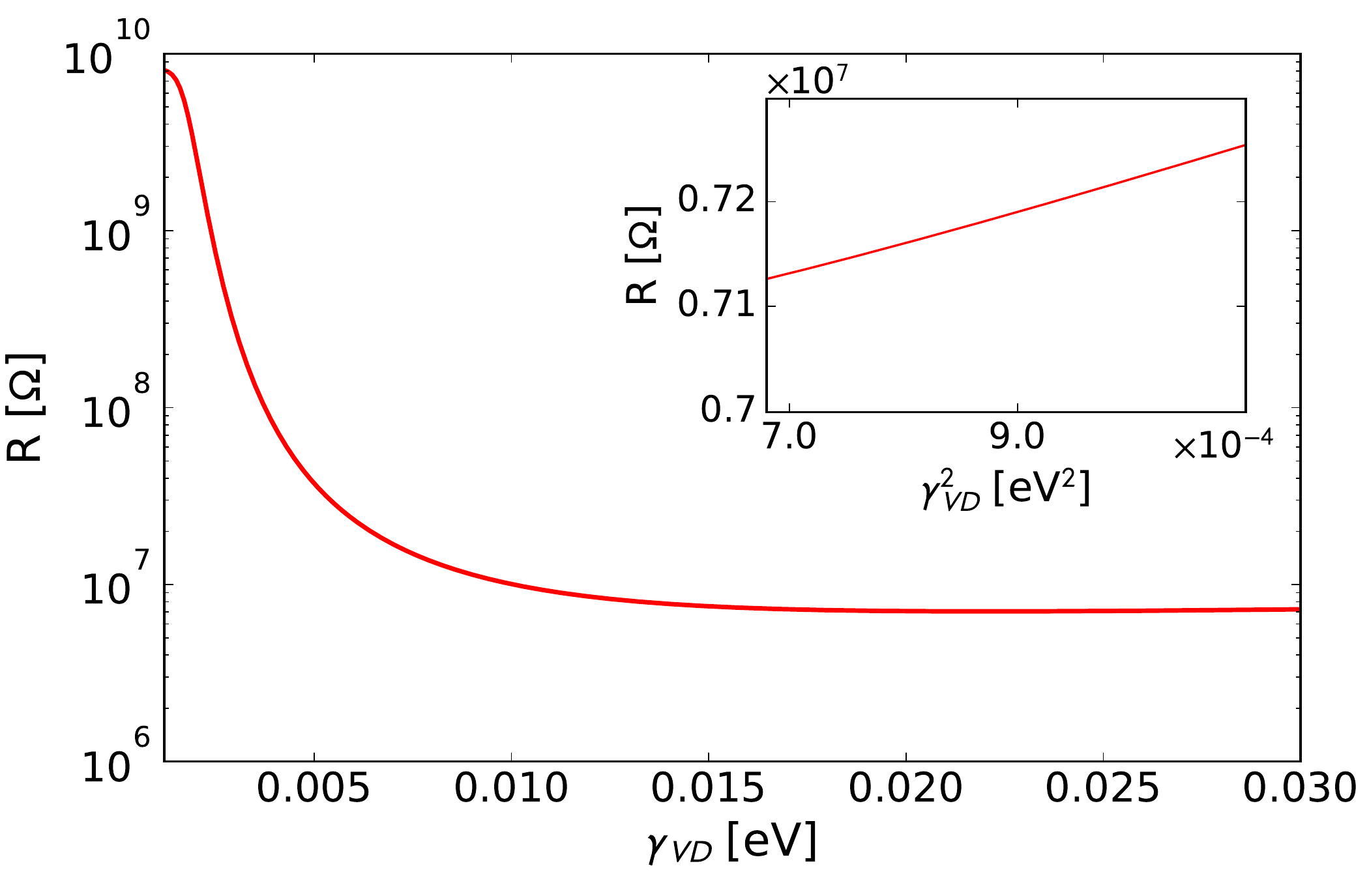}
  \caption{Dephasing (and thus temperature) dependent resistance for a stacked DNA molecule with $n = 8$ base pairs, calculated using the full ladder model. Inset: Zoom in to high temperature regime.}
  \label{Temp_Dependence}
\end{figure}

To get a more detailed view on the influence of dephasing we calculate the resistance of single DNA
molecules dependent on the dephasing strength. The result for the stacked DNA molecule with $n = 8$
base pairs is shown in Fig. \ref{Temp_Dependence}.

In this calculation the dephasing strength was varied proportional to the square root of the
temperature according to the formula (\ref{gamma_vd}).  Analyzing this figure, two distinct
temperature dependencies can be found. At low temperatures the resistance shows a strong variation,
which is the result of level broadening in the case when the resonant transition window is shifted
slightly from the Fermi energy, as is shown in Fig.\,\ref{Transmissions}c.

\begin{figure}[t]
  \includegraphics[width = 0.99\columnwidth]{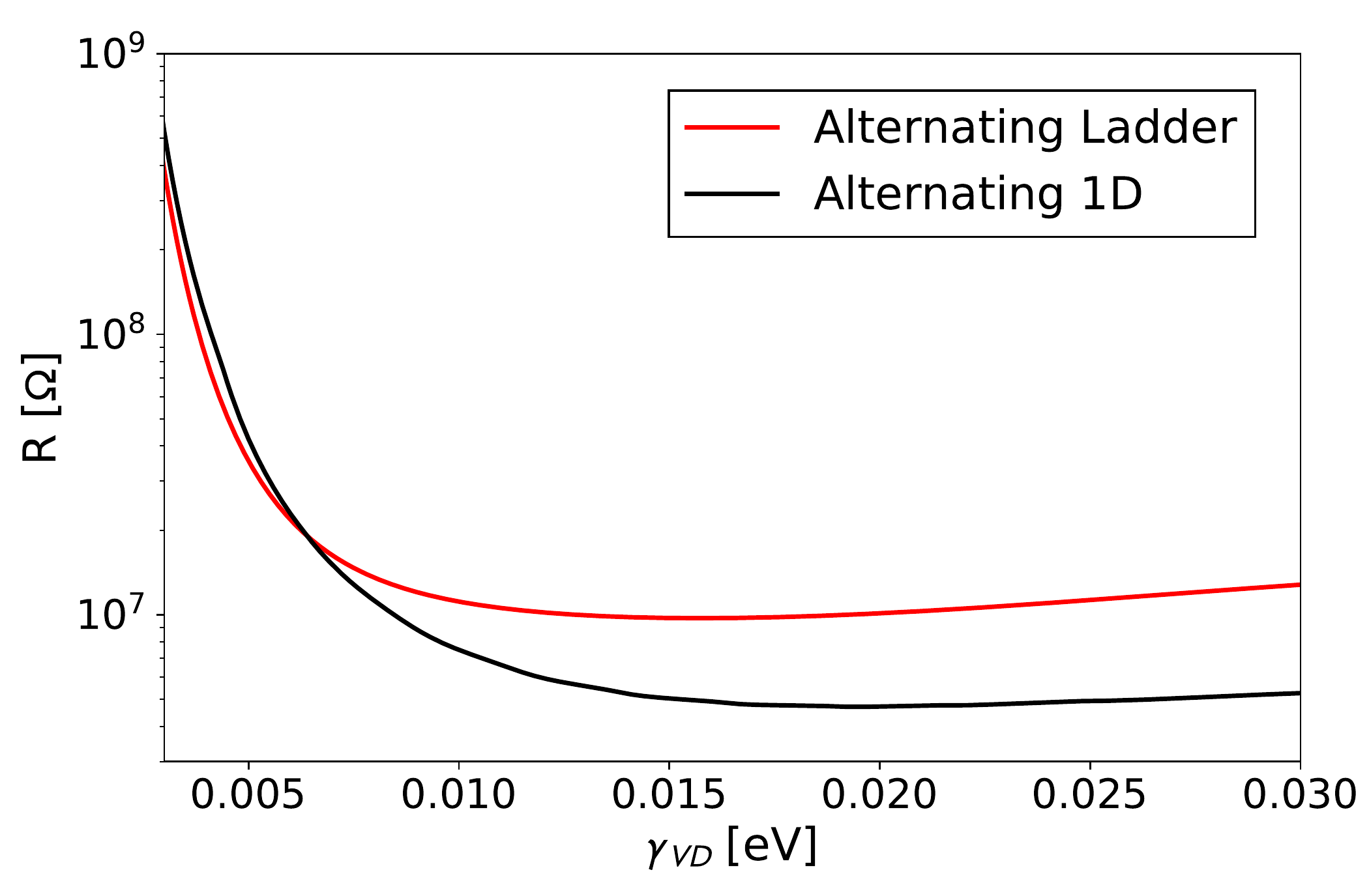}
  \caption{Dephasing (and thus temperature) dependent resistance for an alternating DNA molecule with $n = 8$ base pairs, calculated using the full ladder and 1D models. }
  \label{Alternating_DNA_Models}
\end{figure}

On the other hand for higher temperatures the resistance seems to be constant but if we zoom in we
see a slight increase (see inset in Fig. \ref{Temp_Dependence}). Moreover, the resistance is linear
dependent on $\gamma_{VD}^2$, thus linear in temperature. Such behavior is consistent with
``metallic'' ohmic behavior of the resistance as a function of length. The crossover between the low
temperature and high temperature regimes can be understood with the spectral current densities (see
Fig.\,\ref{Transmissions}). As mentioned before decoherence effects cause broadening as well as
decrease of spectral current density.  For low voltages the current is obtained by the integration
of spectral current density in the small voltage interval. In the low dephasing regime the
broadening of spectral current density dominates resulting in an increasing current and thus
decreasing resistance. On the contrary, the metallic-like regime is mainly affected by the decrease
of spectral current density since the broadening appears outside the integration window and
therefore we obtain a slightly increasing resistance.  Similar results were obtained also for odd
number of base pairs as well as for the alternating DNA molecule.

For alternating case there is considerable difference between ladder and 1D models, which we saw
already in Fig.\,\ref{Tunneling_Models}. From the temperature dependence
(Fig.\,\ref{Alternating_DNA_Models}) it follows that this difference is larger at larger
dephasing. The reason may be that the effects of interference are more essential for the ladder model.

\section{Conclusion}
\label{sec_conclusion}

We have shown that decoherence effects decrease and broad the spectral current density. This
broadening is slightly different in the investigated dephasing models. For alternating DNA molecules
this closes the transmission gaps at Fermi energy occuring in the completely coherent case, leading
to a conducting behavior and causes a monotonous length dependent resistance. On the contrary,
dephasing in stacked DNA structures just closes the transmission gaps for molecules with odd number
of base pairs and therefore molecules with even number of base pairs show
a lower conductance. This distinct transport mechanism were measured by Xiang et al \cite{xiang_intermediate_2015}.\\
Furthermore we have shown that vibronic dephasing leads to crossover from coherent tunneling regime
at low temperature to a metallic-like behavior for higher temperatures. To understand better this
crossover and low temperature transport in DNA, the conductance measurements at low temperature
would be very desired. We also suggest to investigate the interplay between vibronic and conducting
environment dephasing.

\section{Acknowledgement}

We thank B\'alint Aradi and Alessandro Pecchia for valuable discussions and help with the development of the
{\textsc{DFTB$^{\text{+}}$XT\ }} package. This work was supported by the Deutsche
Forschungsgemeinschaft (FR2833/50-1, GRK 2247) and the European Graphene Flagship.

\bibliography{}

\newpage
\appendix
\section{Effective 1D models for DNA}

In section 3 we presented the DNA ladder and the corresponding 1D models, which were compared in
section 4. The 1D alternating DNA molecule can be described by:
\begin{align}
 \boldsymbol{\hat{H}}^\text{A} = \sum_{i=2}^{2n+1} \varepsilon_\text{\tiny G} \hat{c}_i^\dagger  \hat{c}_i + 
		    \sum_{i=1, 2n+2} \varepsilon_\text{\tiny T} \hat{c}_i^\dagger \hat{c}_i + \notag \\ 
		    \sum_{i=1}^{n} t_{3'-\text{\tiny GG}-3'} \left(\hat{c}_{2i}^\dagger \hat{c}_{2i+1} + \hat{c}_{2i+1}^\dagger \hat{c}_{2i} \right) + \notag \\
		    \sum_{i=1}^{n-1} t_{5'-\text{\tiny GG}-5'} \left(\hat{c}_{2i+1}^\dagger \hat{c}_{2i+2} + \hat{c}_{2i+2}^\dagger \hat{c}_{2i+1}  \right) + \notag \\
		    \sum_{i=1,2n+1} t_{3'-\text{\tiny TG}-5'} \left(\hat{c}_{i}^\dagger \hat{c}_{i+1} + \hat{c}_{i+1}^\dagger \hat{c}_{i}  \right) \,,
\end{align}
and the stacked DNA reads:
\begin{align}
\boldsymbol{\hat{H}}^\text{S} = \sum_{i=2}^{2n+1} \varepsilon_\text{\tiny G} \hat{c}_i^\dagger  \hat{c}_i + 
		    \sum_{i=1, 2n+2} \varepsilon_\text{\tiny T} \hat{c}_i^\dagger \hat{c}_i + \notag \\
  		    \sum_{i=2}^{n} t_{3'-\text{\tiny GG}-5'} \left(\hat{c}_{i}^\dagger \hat{c}_{i+1} + \hat{c}_{i+1}^\dagger \hat{c}_{i} \right) + \notag \\
		    \sum_{i=n+2}^{2n} t_{5'-\text{\tiny GG}-3'} \left(\hat{c}_{i}^\dagger \hat{c}_{i+1} + \hat{c}_{i+1}^\dagger \hat{c}_{i} \right) + \notag \\
    		    \sum_{i=1,2n+1} t_{3'-\text{\tiny TG}-5'} \left(\hat{c}_{i}^\dagger \hat{c}_{i+1} + \hat{c}_{i+1}^\dagger \hat{c}_{i}  \right) + \notag \\
    		    t_{3'-\text{\tiny GG}-3'} \left(\hat{c}_{n+1}^\dagger \hat{c}_{n+2} + \hat{c}_{n+2}^\dagger \hat{c}_{n+1} \right) \,.
\end{align}

\section{Charge transfer integrals and onsite energies}
\label{params}

For the sake of completeness we show the used charge transfer integrals as well as the averaged onsite energies of Senthilkumar et al. \cite{senthilkumar_absolute_2005}

\begin{table}[H] 
\centering
\caption{Charge transfer integrals in eV recompiled from \cite{senthilkumar_absolute_2005} \label{table:CT}}
\begin{tabular}{c||cccc}
&&$t_{5'-\text{XY}-3'} = $ &$t_{3'-\text{YX}-5'}$& \\ \hline\hline
\textbf{X} &  & \textbf{Y}  & &  \\ \hline
 &  G & A & C & T \\ \hline
G &  0.053  & -0.077& -0.114& 0.141 \\ 
A & -0.010 & -0.004 & 0.042 & -0.063 \\ 
C & 0.009 & -0.002 & 0.022 & -0.055 \\ 
T & 0.018 & -0.031 & -0.028 & 0.072 \\ \hline
&  & $t_{5'-\text{XY}-5'}$  & &  \\ \hline \hline
 &  G & A & C & T \\ \hline
G &  0.012 & -0.013 &  0.002  & -0.009 \\ 
A & -0.013 &  0.031 & -0.001  & 0.007 \\ 
C &  0.002 & -0.001 &  0.001  & 0.0003 \\ 
T & -0.009 &  0.007 &  0.0003 & 0.001 \\ \hline
 &  & $t_{3'-\text{XY}-3'}$  & &  \\ \hline \hline
 &  G & A & C & T \\ \hline
G & -0.032 & -0.011 & 0.022 & -0.014 \\ 
A & -0.011 &  0.049 & 0.017 & -0.007 \\ 
C &  0.022 &  0.017 & 0.010 &  0.004 \\ 
T & -0.014 & -0.007 & 0.004 &  0.006 \\ \hline 
\end{tabular}
\end{table}

\begin{table}[H] 
\centering
\caption{Averaged site energies and charge transfer integrals within the Watson-Crick base pairs in eV \label{table:site}}
\begin{tabular}{cccccc}
$\varepsilon_\text{\tiny G}$ & $\varepsilon_\text{\tiny C}$ & $\varepsilon_\text{\tiny A}$ & $\varepsilon_\text{\tiny T}$&$t_{\text{GC}}$ &$t_{\text{AT}}$ \\ \hline
8.178 & 9.722 & 8.631 & 9.464&-0.055&-0.047
\end{tabular}
\end{table}

\section{Modification of CT integrals}
\label{mod}

As mentioned before we adjusted the DFT charge transfer integrals from Senthilkumar et
al. \cite{senthilkumar_absolute_2005} to reproduce experimental results obtained by Xiang et
al. \cite{xiang_intermediate_2015}. For low temperature calculations the original parameter set
reproduces the measured behavior quite well. For room temperature the calculations do not show any
significant oscillations, which were found in the experiments, for the stacked DNA molecule.  It is
well known, that the oscillations are caused by partially delocalisation of charge carriers within
the stacked G-sequences \cite{kim_intermediate_2016}. That is why they can be controlled by
$t_{3-\text{GG}-5}$, i.e. the intrastrand charge transfer integral between neighboring G bases. For
our calculations we increased $t_{3-\text{GG}-5}$ from $0.053\,\text{eV}$ to $0.080\,\text{eV}$.
This can be possibly explained by the fact that DFT calculations were made at zero temperature
ground state and the effects of vibrations (geometry changes) at high temperatures are not taken
into account. \\

\end{document}